\documentclass[apj]{emulateapj}
\usepackage{apjfonts}
\usepackage{lscape}

\newcommand{\msun}{$M_{\odot}$}

\shorttitle{Widest VLM Binary}
\shortauthors{Radigan et al.}

\begin{document}

\title{Discovery of the Widest Very Low Mass Field Binary}

\author{Jacqueline Radigan\altaffilmark{1}, David
  Lafreni\`ere\altaffilmark{1}, Ray Jayawardhana\altaffilmark{1}, Ren\'e Doyon\altaffilmark{2}}
\altaffiltext{1}{Department of Astronomy and Astrophysics, University
  of Toronto, 50 St. George Street, Toronto, ON, M5S 3H4, Canada}
\altaffiltext{2}{D\'epartement de Physique and Observatoire du Mont M\'egantic,
  Universit\'e de Montr\'eal, C.P. 6128, Succ. Centre-Ville,
  Montr\'eal, QC, H3C 3J7, Canada}
\email{radigan@astro.utoronto.ca}
\begin{abstract}
We present the discovery of the widest ($\sim6700$~AU) very low mass
field binary to date, found in a proper motion cross-match of the Sloan Digital Sky Survey and the Two Micron All Sky Survey.  Our follow-up J-band imaging provides a 10-year baseline for measuring proper motions.  Consequently, we are able to confirm the common proper motion of the pair to within 10~mas, implying a 99.5\% probability of their physical association.  
Near infrared spectra of the components indicate spectral
types of M6$\pm1$ and M7$\pm1$.  The system resides at a spectroscopic distance of
$105\pm13$~pc and has an angular separation of $63.38\pm 0.05\arcsec$.  We have
used evolutionary models to infer component masses of
$0.105^{+0.029}_{-0.017} $~\msun~ and
$0.091^{+0.010}_{-0.007}$~\msun.~  The large separation and low binding energy of this
system can provide constraints for formation models of very low mass stars.     
\end{abstract}

\section{Introduction}
Studies of the low end of the stellar initial mass function have made great
strides in recent years, and have revealed a large population of very low
mass (VLM, ~M $\lesssim$0.1~\msun) stars and brown dwarfs (BDs) in the Galaxy. Their ubiquity poses
a challenge to traditional models of star formation via gravitational
fragmentation, because their masses are well below the typical Jeans mass in
molecular clouds. Whether these cool objects form similarly to more massive
stars or require additional processes is intensely debated
\citep[e.g.][]{burgasser07,luhman07,whitworth07}.  Since multiplicity properties
can provide insight into formation scenarios, it is of considerable
interest to determine whether the properties of VLM binaries differ from
those of more massive stars.  Trends pertaining to VLM binaries are
discussed at length by \citet{burgasser07}.  These systems are observed to be
less frequent, more tightly bound, and of higher mass ratios than
their more massive counterparts.   It is not clear whether these
properties form part of a continuous mass dependent trend, or
represent a unique population with a potentially different formation mechanism.

In recent years much attention has been paid to relatively rare wide ($>$ 100 AU) VLM
binaries since their large separations and low binding energies provide direct constraints
for formation models.  In particular, their existence challenges
the ejection model for the formation of VLM stars and BDs
\citep{reipurth01} since such fragile systems are not expected to
survive the ejection process.

To date there are 8 known widely separated VLM and BD
binaries, both in low-density star forming regions
\citep[e.g.][]{luhman04,jayawardhana06,allers06,close07,luhman09} and in the field
\citep[e.g.][]{billeres05,caballero07a,artigau08,gizis00,bejar08}.  
Surveys indicate a wide companion fraction for VLM stars and BDs of no
more than 1-2\% \citep{burgasser07,caballero07b,allen08}. However,
searches for ultracool binaries at the largest separations remain
incomplete.  The recent discovery of the widest VLM binary by \citet{artigau07}
with a separation of $\sim5100$~AU, and the discovery of a
$\sim1800$~AU binary by \citet{caballero07a} point to the existence of
a population of such ultrawide systems.

Here we report the discovery of the widest VLM
binary system to date with a projected separation of $\sim6700$~AU.

\section{Discovery and Observations}\label{sect:discovery}

\begin{deluxetable*}{llllll}[float]
\tablewidth{7.0in}
\tabletypesize{\small}
\tablecolumns{6}
\tablecaption{Multiple epochs of imaging observations for 2M1258AB\label{tbl:tab1}}
\tablehead{
\colhead{Source} & \colhead{Epoch (yr) } & \colhead{$\alpha$
  $A$ (deg)} &
\colhead{$\alpha$ $B$ (deg)} & \colhead{$\delta$ $A$ (deg)} &
\colhead{$\delta$ $B$ (deg)}\\
}
\startdata
  2MASS & 1998.2666 & 194.645879(0.06) &  40.218994(0.08) & 194.658278(0.06) &  40.233826(0.08)\\
  SDSS & 2003.3149 & 194.645997(0.05) &  40.218824(0.04) & 194.658460(0.06) &  40.233639(0.06)\\
  CPAPIR  & 2008.2292 & 194.646157(0.08) &  40.218670(0.08) &  194.658563(0.08) &  40.233519(0.08)\\
  CPAPIR & 2008.4999 & 194.646156(0.08) &  40.218668(0.08) &  194.658549(0.08) &  40.233522(0.08)\\
\enddata
\tablecomments{SDSS and 2MASS positions are taken directly from their respective catalogs.  The uncertainties are given in the brackets, in units of arcseconds.}
\end{deluxetable*}

The system, comprised of 2MASS J12583501+4013083 and 2MASS J12583798+4014017 (2M1258AB hereafter), was found in a cross-match of the Sloan Digital Sky Survey
(SDSS) 6th data release Photoprimary Catalog
\citep{sdss} and the Two Micron All Sky Survey (2MASS) Point Source Catalog \citep{skrutskie06}
in which we searched for common proper motion pairs containing VLM
components.  The correlation of catalogs, calculation of proper motions, and the
identification of co-moving stars was done in overlapping sections of 4~$\rm{deg^2}$
of the sky at a time, spanning the entire contiguous region of the
SDSS Legacy survey in the northern galactic cap.  For every 2MASS
source the closest SDSS match was found and proper motion vectors with
uncertainties were computed.  A cut was made in order to select only
stars that had moved at the $3\sigma$ level compared to all other stars within the area.
Stars within 120\arcsec~of one another with proper motion amplitudes
agreeing within 2$\sigma$ and proper motion components agreeing within
1$\sigma$ in one of right ascension  and declination, were flagged as potential binaries.
The pair 2M1258AB was found by applying further color cuts of $z^{\prime}-J
>1.5$, corresponding approximately to mid-M spectral types and later, to
both components of potential binaries within our subsample.  

To determine spectral types for the components of 2M1258AB we obtained
near-infrared spectra (R$\sim$250) on 2008 March 01 using the SpeX Medium-Resolution Spectrograph \citep{spex} at NASA's Infrared
Telescope Facility (IRTF). Observations were made in prism mode with
the $0.3\arcsec$ slit. We obtained six 120~s exposures arranged in
three AB cycles.  For telluric and instrumental transmission correction, the A0 star HD~109615 was
observed immediately after the target at the same airmass.  The
spectra were reduced using SpeXtool
\citep{spextool1,spextool2}.

To confirm the common proper motion of this system, follow-up $J$-band imaging observations were obtained on 2008 March 25 and
2008 July 01 with the wide field near-infrared camera {\em CPAPIR} (E.
Artigau et al., in preparation) at the 1.6-m Mont-M\'egantic
telescope.  At each visit, 16 exposures of 8.12~s were obtained by dithering the telescope
by $\sim$5\arcsec\ between exposures.  The follow-up $J$-band imaging data were reduced as follows.  A sky frame was first constructed by taking the
median of all images after masking the stars in each image. After
subtraction of this sky frame, the images were divided by a normalized flat
field image. The distortion and astrometry solutions of each image were
computed by using the 2MASS PSC as a reference frame.  The reduced and
calibrated images were then combined by taking their median.  Astrometric uncertainties of $\sim 80$~mas were determined from the dispersion of
the positions of all stars in the field with respect to their positions in the 2MASS PSC. The SDSS astrometry used was taken directly from the SDSS catalog.  The positions reported by the catalog are derived from the $r^{\prime}$ band in which our sources are very faint.  Therefore, as an additional check we calculated SDSS positions directly from the $i^\prime$ and $z^\prime$ images, and found them to be consistent with the catalog positions within the quoted uncertainties. A list of all astrometric measurements from OMM, SDSS and 2MASS used to compute proper motions are listed
in table \ref{tbl:tab1}.

Proper motions and their uncertainties were obtained using
an error-weighted linear regression, yielding proper motions for the
primary and secondary respectively of $76\pm8~{\rm mas~yr^{-1}}$ and $76\pm 8~{\rm mas~yr^{-1}}$ in right ascension,
and $-115\pm9~{\rm mas~yr^{-1}}$ and $-108\pm 10~{\rm mas~yr^{-1}}$ in
declination.  The proper motions are in agreement within the $1\sigma$
errors.  The mean angular separation of the system was found to be
$63.38\pm0.05\arcsec$. 

Using the space density for M dwarfs given by \citet{phanbao03} and our total
search area of 7668 ${\rm deg^2}$, we would expect to find 1.8 pairs of M6-M8 dwarfs within $64\arcsec$ of one
another and at distances ranging from 75-125~pc (spectroscopic distances are
derived in \S\ref{sect:properties}) within our search area.  Using
the space velocity distributions from \citet{bochanski05} the probability of any two M
dwarfs having proper motions greater than $0.1\arcsec~{\rm yr^{-1}}$ and agreeing within
twice our uncertainties is 0.003.  Therefore, we expect to find at
most 0.0054 unrelated pairs similar to 2M1258AB in our entire search area,
which implies a 99.5\% probability of physical association for
2M1258AB.

For thoroughness, we note that the angular separation of 2M1258AB appears to increase monotonically over
the three imaging epochs (2MASS, SDSS, CPAPIR) based on positions provided in table 1.  However, the trend falls within our uncertainties and is completely consistent with a constant separation over time.  Furthermore, upon examination of the 2MASS ATLAS images, we find that this apparent trend is broken when measuring the angular separation directly from the $J$ and $H$ band images, and only holds using the $K_s$ band image.  Therefore, the trend appears to be coincidental, and it remains most likely that the system is bound.

\section{Physical Properties of the Components}\label{sect:properties}
A summary of observational and physical properties of 2M1258AB is given in table \ref{tbl:tab2}. 

\begin{figure}[here]
\epsscale{1}
\plotone{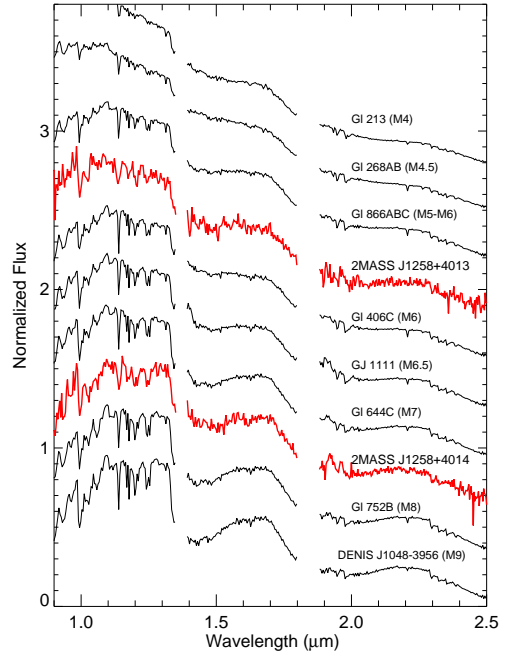}
\caption{Spectra for 2M1258AB taken with the SpeX prism at the IRTF,
  alongside comparison spectra of field dwarfs from the IRTF spectral library \citep{cushing05}.
\label{fig:spectra}}
\end{figure}

Spectral types for the components were determined primarily from the best-fitting reference spectra from the IRTF
spectral library \citep{cushing05}.  Reasonable fits were found from M5\footnote{the M5 match was to Gl 866ABC, which has an
  alternate classification of M6 by \citet{reid04}}-M6.5 for the
primary, and M6.5-M8 for the secondary.  Figure \ref{fig:spectra} shows our
spectra of 2M1258AB plotted alongside those of other M4-M9 dwarfs for comparison. 

A weighted average of the H$_2$O A, B, and C indices\footnote{we have
  chosen no to include the H$_2$O D  index as it predicts unreasonably
  early spectral types for both components, possibly a systematic effect of the telluric
correction} defined by \citet{mclean03} indicates spectral types of M6$\pm0.8$ and M7$\pm0.8$
for the primary and secondary respectively, in agreement with the
former determination.  Based on our fits to NIR reference spectra and the H$_2$O spectral indices we
assign spectral types of M6$\pm1$ for the primary and M7$\pm1$ for the
secondary.  The error bars are representative of the
ambiguity present in assigning M dwarf spectral types based on low
resolution NIR spectra, as is evident from figure \ref{fig:spectra}.

In order to determine absolute $J$, $H$, and, $K_s$ magnitudes and
their associated uncertainties we compiled a list of M dwarfs with
measured parallaxes \citep{ctiopi1,ctiopi2,ctiopi3}.  From this list, M dwarfs with spectral types within 0.5
subclass of the component spectral types were used to determine mean
absolute magnitudes for 2M1258AB, and the standard deviation of M
dwarfs within 1.0 subclass was used as a measure of the uncertainty.
By comparing our derived absolute magnitudes to the 2MASS $J$, $H$,
and $K_s$ magnitudes we calculated spectroscopic distances for the
components of $115\pm20$~pc and $95\pm18$~pc for the primary and
secondary respectively.  This implies a projected separation
of $6700\pm800$~AU at the average system distance of $105\pm13$~pc.  

The equivalent widths (EWs) of the 1.25~${\rm \mu m}$ \ion{K}{1} and 1.20~${\rm \mu m}$ FeH features can be used as rough gravity indicators in NIR spectra.  Using the wavelength regions defined by \citet{gorlova03} we measured the \ion{K}{1} ~EWs to be $5.0\pm1.7$, $10.6\pm1.6$  and the FeH ~EWs to be $12.9\pm3.5$, $6.6\pm3.2$ for the primary and the secondary respectively.  When compared with figure 8 of \citet{gorlova03}, the measured EWs--with the exception of the FeH EW in the secondary's spectrum--are most consistent with that of field dwarfs with $\log g > 4.5$, and provide no evidence of youth.  However, the large errors due to the low S/N of our spectra make it impossible to draw firm conclusions about the age or gravity of 2M1258AB. Assuming an age of 1-5~Gyr (as an upper limit, see \S\ref{sect:discussion}) the component masses can be estimated from
evolutionary models \citep{baraffe98}.  Masses of
$0.105^{+0.029}_{-0.017}$~\msun~ and $0.091^{+0.01}_{-0.007}$~\msun~ were found
for the primary and secondary respectively by minimizing the
error-weighted sum of square deviations between our derived absolute
$JHK_s$ magnitudes and the models, with errors bars representing $1\sigma$
deviations from the minimum.  This determination is also consistent with
masses derived from effective temperatures rather than magnitudes.
The mass ratio of 2M1258AB is $0.87\pm0.25$, consistent with other
wide VLM binaries which tend to have mass ratios close to unity \citep{burgasser07}.

\begin{deluxetable}{lcc}
\tablewidth{0pt}
\tabletypesize{\small}
\tablecolumns{15}
\tablecaption{Physical Properties of 2M1258AB\label{tbl:tab2}}
\tablehead{
\colhead{Quantity} & \colhead{A} & \colhead{B}
}
\startdata
2MASS Designation & J12583501+4013083 & J12583798+4014017\\
$\mu_{\alpha}\cos{\delta}$ (mas~${\rm yr^{-1}}$) &  $76\pm8$      &  $76\pm8$ \\
$\mu_{\delta}$ (mas~${\rm yr^{-1}}$)       &  $-115\pm9$        &  $-108\pm10$ \\
$J$  (mag)\tablenotemark{a}         &  $15.59\pm0.05$         &  $15.61\pm0.05$ \\
$H$  (mag)\tablenotemark{a}          &  $14.84\pm0.06$         &  $14.89\pm0.06$ \\
$K_s$  (mag)\tablenotemark{a}          &  $14.43\pm0.06$         &  $14.62\pm0.07$ \\
$r^{\prime}$ (mag) \tablenotemark{b} &  $20.74\pm0.04$ & $22.00\pm0.07$ \\
$i^{\prime}$ (mag) \tablenotemark{b} & $18.40\pm0.02$ & $19.18\pm0.03$\\
$z^{\prime}$ (mag) \tablenotemark{b} & $17.18\pm0.02$ & $17.68\pm0.02$\\
$M_J$   (mag)\tablenotemark{c}       &  $10.20\pm0.70$         &  $10.71\pm0.41$ \\
$M_H$   (mag)\tablenotemark{c}       &  $9.55\pm0.64$          &  $10.07\pm0.40$ \\
$M_{Ks}$(mag)\tablenotemark{c}       &  $9.17\pm0.62$          &  $9.70\pm0.37$  \\
d (pc)              &  $115\pm20$             &  $95\pm18$      \\
Spectral type       &  M6$\pm1$               &  M7$\pm1$       \\
$T_{eff}$ (K)\tablenotemark{d}       &  $2850\pm300$           & $2620\pm170$   \\
Mass (\msun)        &  $0.105^{+0.029}_{-0.017}$         & $0.091^{+0.01}_{-0.007}$  \\
Angular separation ($\arcsec$)& \multicolumn{2}{c}{$63.38\pm0.05$} \\
Projected separation (AU) & \multicolumn{2}{c}{$6700\pm800$} \\
Binding energy ($10^{41}$~erg) & \multicolumn{2}{c}{$0.25\pm0.08$} \\
Mass ratio & \multicolumn{2}{c}{$0.87\pm0.25$} \\
\enddata
\tablenotetext{a}{magnitudes from the 2MASS Point Source Catalog}
\tablenotetext{b}{PSF magnitudes from the SDSS DR6 Catalog}
\tablenotetext{c}{Estimates based on M dwarfs within $\pm1$ spectral
  type with measured parallaxes}
\tablenotetext{d}{based on the $T_{eff}$ versus spectral type relationship of \citet{golimowski04}}
\end{deluxetable}

\section{Discussion}\label{sect:discussion}

\begin{figure*}[float]
\epsscale{1}
\plottwo{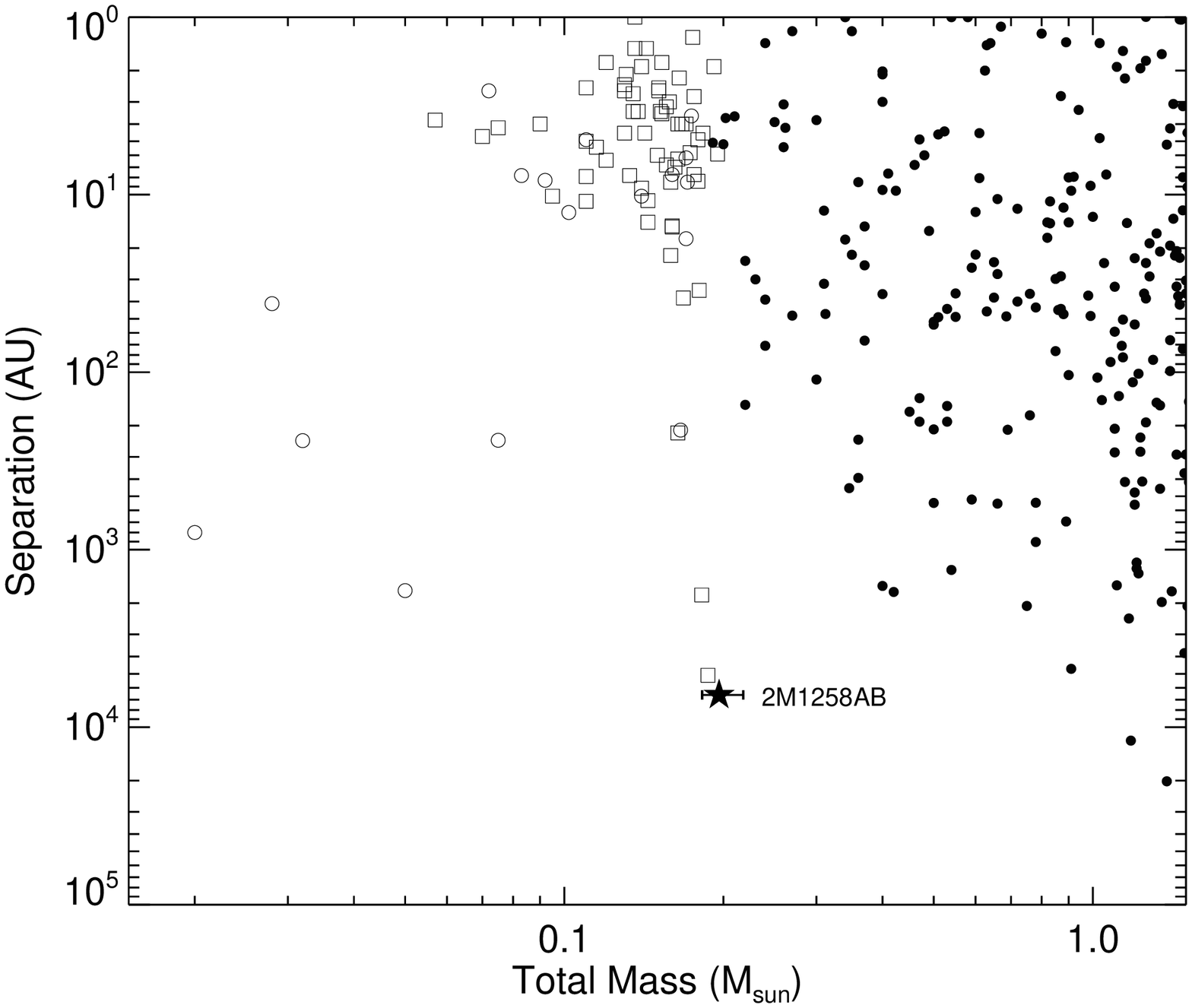}{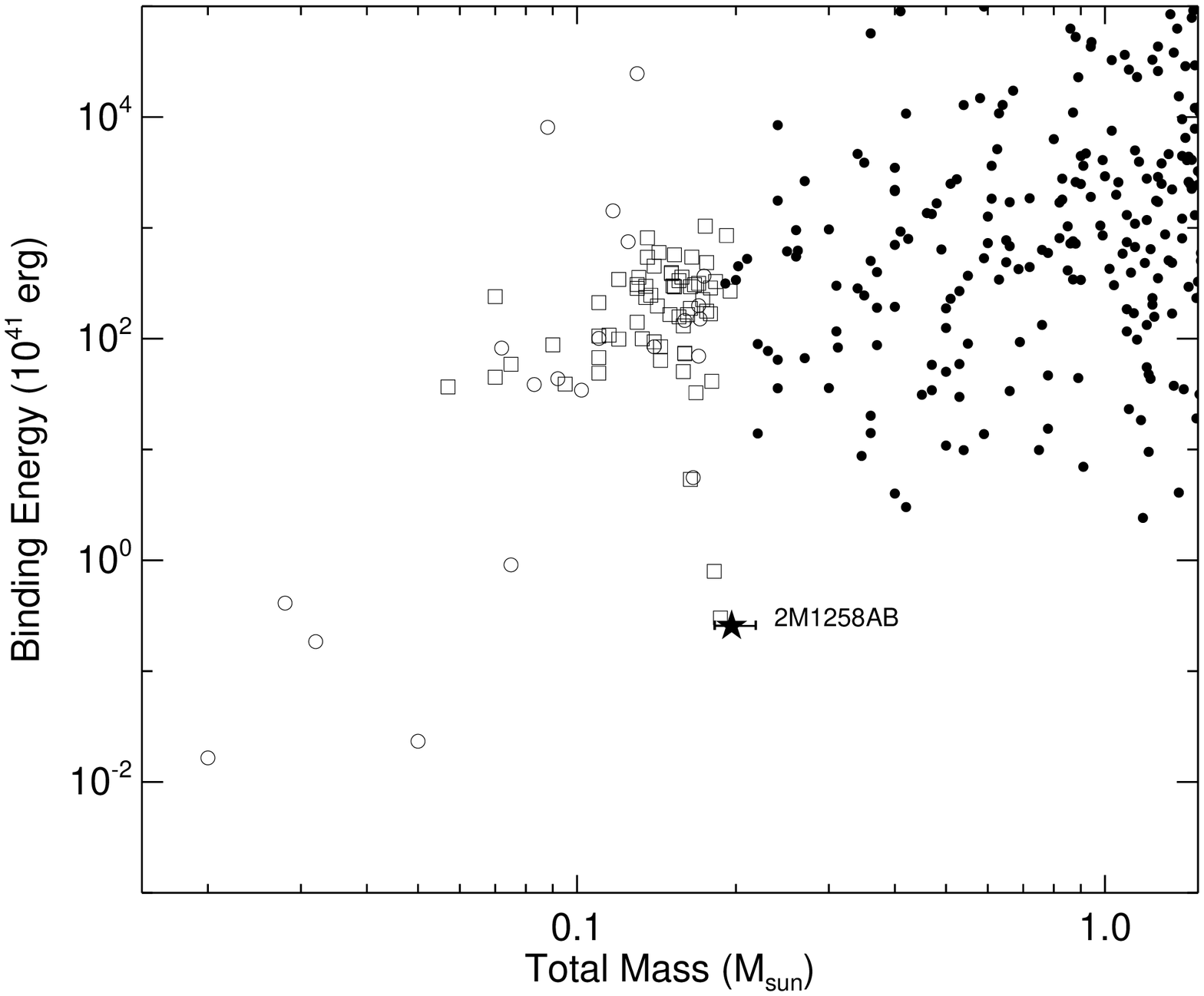}
\caption{Separation and binding energy versus total system mass for known binary systems.
  Stellar binaries (dots) are from \citet{tokovinin97,FM92,reid97,close90,kraus07b}.  VLM
  binaries (open squares, open circles) are from \citet{burgasser07,caballero06,luhman09} and the VLM binary archive, maintained by Nick
  Siegler at {\em http://www.vlmbinaries.org}.  Open circles identify VLM binaries that are known members of young associations.
\label{fig:sep}}
\end{figure*}

Of the previously known VLM binaries, only 2 have separations $>
1000$~AU: K\"o1~AB with a separation of $\sim1800$~AU
\citep{caballero07b} and 2M0126AB with a separation of $\sim5100$~AU
\citep{artigau07}.  In addition, a probable VLM binary with a separation of $\sim1700$~AU has been identified by \citet{caballero06} in the $\sigma$ Orionis cluster.  With a separation of $\sim6700$~AU 2M1258AB is the widest VLM binary yet discovered (see figure
\ref{fig:sep}).  The existence of 3-4 of such ultrawide systems is strong evidence that they are not
statistically rare chance alignments, but rather natural outcomes of star
formation. 

Many other authors have discussed the implications of wide VLM
binaries for various formation models \citep[e.g.][]{luhman04,artigau07}.
As with other wide VLM binaries, 2M1258AB would not have survived strong
dynamical interactions during the formation process and is
inconsistent with ejection models \citep{reipurth01}.  Formation of
the secondary in the disk of the primary
\citep[e.g.][]{stamatellos07,boss00} can also be strongly ruled out
since disks around low mass stars tend to be at most a few hundred AU in diameter \citep{vicente05}, and tend to be only a few percent as massive
as the stars they surround \citep{scholz06}.  Thus this mechanism is unlikely to
produce near-equal mass binaries, nor binaries with such wide
separations.  Alternatively, since it is not uncommon for
higher mass stars to form weakly bound binaries \citep[e.g.][]{duquennoy91},
models in which VLM stars form similarly to their more massive
counterparts are better able to account for the existence of wide VLM
pairs.  Turbulent fragmentation \citep{padoan02} is one such model,
capable of producing low-mass cores down to $\sim3$ Jupiter
masses. 

Higher order multiplicity may result in higher total
system masses for wide VLM binaries, implying larger Jeans masses in
formation.  For more massive binaries, a large fraction
($\sim25$\%) of wide ($>1000$~AU) binaries are higher order multiples
\citep{makarov08}.  Of the two previously known ultrawide VLM
binaries, at least one, K\"o1~AB, is a candidate triple system, where
the secondary is a suspected spectroscopic double with total mass of $\sim0.08$~\msun
 ~\citep{basri06}. Other known examples among low-mass and VLM stars
include: LP~213-68A(BC) with M6.5 and M8 components and a separation
of 230~AU \citep{gizis00,close03}; G~124-62ABab, M4.5/L1, 1900~AU
\citep{seifahrt05};  GJ 1245ABC,(M5.5/M5.5)/M8, 38~AU
\citep{mccarthy88}; USco J160611.9-193532 AB, M5/M5,1600~AU \citep{kraus07b}; and candidate triple LP
714-37ABC M5.5/(M7/M8), 33~AU \citep{phanbao06}.  Based on 2MASS
and SDSS photometry there is some evidence that 2M1258AB may also be a triple system.  
%We note
%that the primary is clearly brighter than the secondary in the
%$i^\prime$ and $z^\prime$ bands, however the components have almost identical J and H magnitudes, suggesting that
%the secondary may itself be an unresolved binary.  
Although the primary clearly has an earlier spectral type than the secondary (see figure \ref{fig:spectra}), the apparent $J$, $H$, and $K$ magnitudes of the components are nearly indentical.  Referring to table 3 of \citet{hawley02}, we note that the difference in mean absolute J magnitude between M6 and M7 dwarfs is $\sim0.74$~mag, suggesting that the secondary may itself be an unresolved equal-mass binary.
If both K\"o1AB and 2M1258AB
turn out to be triples then this would suggest that higher
order multiplicity is common among the widest VLM systems.  The
second widest VLM binary, 2M0126AB, has yet to be checked for additional
companions.  

Wide weakly bound binaries such as 2M1258AB are subject to orbital
evolution and eventual disruption as they travel through the Galaxy,
due to encounters with stars and giant molecular clouds.  Orbital evolution is an important
consideration when assessing the  probability that 2M1258AB formed in such
a wide orbit, and also provides a rough constraint on the system age.  The theoretical framework and numerical results for the
evolution and lifetimes of wide binary systems in the solar neighborhood is provided by
\citet{weinberg87}, and can be easily scaled to apply to VLM systems \citep[e.g.,][]{artigau07}. 
From figure 1a of \citet{weinberg87}, the net effect of
diffusive collisions is that an ensemble of wide systems will retain
approximately the same average separation over time, provided they
have not been disrupted.  The dispersion in separation
after $\sim1$~Gyr is on the order of $\Delta \log{a} \approx \log{2}$.
Thus the orbital separation of wide VLM binaries remains approximately within a factor of 2 of the original
separation.  Even allowing for the possibility that binaries such as 2M1258AB as well as
2M0216AB were formed at half their current separation, they would still have initial separations an order of magnitude
greater than that of other wide VLM binaries and present the same
challenges for formation scenarios.  The disruption lifetimes of wide binaries
are also of interest as they can provide upper limits on the system
age.  An order of magnitude estimate of the survival half-life, using
figure 2 of \citet{weinberg87} for a binary with
$a/M_{tot}=0.16$~pc~\msun$^{-1}$, yields $t_{1/2} \approx 1$~Gyr.
If we assume the physical separation of the system to be greater than
the projected separation then the survival half-life is slightly lower
than this.   Since it is unlikely that 2M1258AB is more than a few
times older than $t_{1/2}$, the upper limit
on the system age of 1-5~Gyr used for deriving masses in \S\ref{sect:properties} is reasonable.
 
Further characterization of this system would benefit from optical
spectra to better constrain the spectral types and  activity of the components, as well as a parallax measurement to obtain a more accurate distance and hence separation.  Both the primary and secondary should be checked for additional companions.

\acknowledgements
We would like to thank the anonymous referee for a thorough and thoughtful review, leading to a much improved manuscript.  We would also like to thank Lison Malo for obtaining follow-up imaging observations of
2M1258AB with {\em CPAPIR} at OMM.  JR is supported in part by a
Canadian Graduate Scholarship from the Natural Sciences and Engineering
Research Council (NSERC), Canada.  DL is
supported in part by a postdoctoral fellowship from the Fonds
Qu\'eb\'ecois de la Recherche sur la Nature et les Technologies.  This work was supported in part
by grants to RJ and RD from NSERC, Canada, and an Early Researcher Award from
the province of Ontario to RJ.


\begin{thebibliography}
{}

\bibitem[Adelman-McCarthy et al.(2008)]{sdss} Adelman-McCarthy, J.~K., \& et
al.\ 2008, VizieR Online Data Catalog, 2282, 0
\bibitem[Allen \& Reid(2008)]{allen08} Allen, P.~R., \& Reid, I.~N.\ 2008, ArXiv e-prints, 804, arXiv:0804.2872
\bibitem[Allers et al.(2006)]{allers06} Allers, K.~N., 
Kessler-Silacci, J.~E., Cieza, L.~A., \& Jaffe, D.~T.\ 2006, \apj, 644, 364
\bibitem[Artigau et al.(2007)]{artigau07} Artigau, {\'E}., 
Lafreni{\`e}re, D., Doyon, R., Albert, L., Nadeau, D., 
\& Robert, J.\ 2007, \apjl, 659, L49
\bibitem[Artigau et al.(2008)]{artigau08} Artigau, E., 
Lafreniere, D., Albert, L., \& Doyon, R.\ 2008, arXiv:0810.4540
\bibitem[Baraffe et al.(1998)]{baraffe98} Baraffe, I., Chabrier, G., Allard, F., \& Hauschildt, P.~H.\ 1998, \aap, 337, 403
\bibitem[Basri \& Reiners(2006)]{basri06} Basri, G., \& Reiners, A.\ 2006, \aj, 132, 663
\bibitem[Bate \& Bonnell(2005)]{bate05} Bate, M.~R., \& Bonnell, I.~A.\ 2005, \mnras, 356, 1201
\bibitem[B{\'e}jar et al.(2008)]{bejar08} B{\'e}jar, V.~J.~S., 
Zapatero Osorio, M.~R., P{\'e}rez-Garrido, A., {\'A}lvarez, C., 
Mart{\'{\i}}n, E.~L., Rebolo, R., Vill{\'o}-P{\'e}rez, I., 
\& D{\'{\i}}az-S{\'a}nchez, A.\ 2008, \apjl, 673, L185
\bibitem[Bill{\`e}res et 
al.(2005)]{billeres05} Bill{\`e}res, M., Delfosse, X., Beuzit, J.-L.,
Forveille, T., Marchal, L., \& Mart{\'{\i}}n, E.~L.\ 2005, \aap, 440,
L55
\bibitem[Bochanski et al.(2005)]{bochanski05} Bochanski, J.~J., 
Hawley, S.~L., Reid, I.~N., Covey, K.~R., West, A.~A., Tinney, C.~G., 
\& Gizis, J.~E.\ 2005, \aj, 130, 1871
\bibitem[Boss(2000)]{boss00} Boss, A.~P.\ 2000, \apjl, 536, 
L101 
\bibitem[Bonnell et al.(2008)]{bate08} Bonnell, I.~A., Clark, 
P.~C., \& Bate, M.~R.\ 2008, ArXiv e-prints, 807, arXiv:0807.0460 
\bibitem[Burgasser et al.(2007)]{burgasser07} Burgasser, A.~J.,Reid, I.~N., Siegler, N., Close, L., Allen, P., Lowrance, P., \& Gizis, J.\
2007a, Protostars and Planets V, 427
\bibitem[Caballero(2007b)]{caballero07b} Caballero, J.~A.\ 2007b, 
\apj, 667, 520
\bibitem[Caballero(2007a)]{caballero07a} Caballero, J.~A.\ 2007a, \aap, 462, L61 
\bibitem[Caballero et 
al.(2006)]{caballero06} Caballero, J.~A., Mart{\'{\i}}n, E.~L., Dobbie, P.~D., \& Barrado Y Navascu{\'e}s, D.\ 2006, \aap, 460, 635 


\bibitem[Costa et al.(2006)]{ctiopi2} Costa, E., M{\'e}ndez, 
R.~A., Jao, W.-C., Henry, T.~J., Subasavage, J.~P., 
\& Ianna, P.~A.\ 2006, \aj, 132, 1234
\bibitem[Close et al.(1990)]{close90} Close, L.~M., Richer, 
H.~B., \& Crabtree, D.~R.\ 1990, \aj, 100, 1968 
\bibitem[Close et al.(2003)]{close03} Close, L.~M., Siegler, N., Freed, M., \& Biller, B.\ 2003, \apj, 587, 407
\bibitem[Close et al.(2007)]{close07} Close, L.~M., et al.\ 
2007, \apj, 660, 1492 
\bibitem[Cushing et al.(2004)]{spextool1} Cushing, M.~C., Vacca, W.~D., \& Rayner, J.~T.\ 2004, \pasp, 116, 362
\bibitem[Cushing et al.(2005)]{cushing05} Cushing, M.~C., Rayner, 
J.~T., \& Vacca, W.~D.\ 2005, \apj, 623, 1115 
\bibitem[Duquennoy 
\& Mayor(1991)]{duquennoy91} Duquennoy, A., \& Mayor, M.\ 1991, \aap,
248, 485
\bibitem[Fischer
\& Marcy(1992)]{FM92} Fischer, D.~A., \& Marcy, G.~W.\ 1992, \apj, 396, 178
\bibitem[Gizis et al.(2000)]{gizis00} Gizis, J.~E., Monet, 
D.~G., Reid, I.~N., Kirkpatrick, J.~D., \& Burgasser, A.~J.\ 2000,
\mnras, 311, 385
\bibitem[Golimowski et al.(2004)]{golimowski04} Golimowski, D.~A., 
et al.\ 2004, \aj, 127, 3516
\bibitem[Gorlova et al.(2003)]{gorlova03} Gorlova, N.~I., Meyer, 
M.~R., Rieke, G.~H., \& Liebert, J.\ 2003, \apj, 593, 1074 
\bibitem[Hawley et al.(2002)]{hawley02} Hawley, S.~L., et al.\ 
2002, \aj, 123, 3409
\bibitem[Henry et al.(2006)]{ctiopi3} Henry, T.~J., Jao, W.-C., 
Subasavage, J.~P., Beaulieu, T.~D., Ianna, P.~A., Costa, E., 
\& M{\'e}ndez, R.~A.\ 2006, \aj, 132, 2360
\bibitem[Jao et al.(2005)]{ctiopi1} Jao, W.-C., Henry, T.~J., 
Subasavage, J.~P., Brown, M.~A., Ianna, P.~A., Bartlett, J.~L., Costa, E., 
\& M{\'e}ndez, R.~A.\ 2005, \aj, 129, 1954 
\bibitem[Jayawardhana 
\& Ivanov(2006)]{jayawardhana06} Jayawardhana, R., \& Ivanov, V.~D.\
2006, Science, 313, 1279
\bibitem[Kratter 
\& Matzner(2006)]{kratter06} Kratter, K.~M., \& Matzner, C.~D.\ 2006,
\mnras, 373, 1563
\bibitem[Kraus 
\& Hillenbrand(2007a)]{kraus07} Kraus, A.~L., \& Hillenbrand, L.~A.\
2007a, \apj, 662, 413 
\bibitem[Kraus 
\& Hillenbrand(2007b)]{kraus07b} Kraus, A.~L., \& Hillenbrand, L.~A.\ 2007b, \apj, 664, 1167 
\bibitem[Luhman et al.(2009)]{luhman09} Luhman, K.~L., Mamajek, 
E.~E., Allen, P.~R., Muench, A.~A., 
\& Finkbeiner, D.~P.\ 2009, \apj, 691, 1265 


\bibitem[Luhman(2004)]{luhman04} Luhman, K.~L.\ 2004, \apj, 614, 
398
\bibitem[Luhman et al.(2007)]{luhman07} Luhman, K.~L., Joergens, 
V., Lada, C., Muzerolle, J., Pascucci, I., 
\& White, R.\ 2007, Protostars and Planets V, 443 
\bibitem[Makarov et al.(2008)]{makarov08} Makarov, V.~V., 
Zacharias, N., \& Hennessy, G.~S.\ 2008, ArXiv e-prints, 808, arXiv:0808.3414 
\bibitem[McCarthy et al.(1988)]{mccarthy88} McCarthy, D.~W., Jr., 
Henry, T.~J., Fleming, T.~A., Saffer, R.~A., Liebert, J., 
\& Christou, J.~C.\ 1988, \apj, 333, 943 
\bibitem[McLean et al.(2003)]{mclean03} McLean, I.~S., McGovern, 
M.~R., Burgasser, A.~J., Kirkpatrick, J.~D., Prato, L., 
\& Kim, S.~S.\ 2003, \apj, 596, 561
\bibitem[Padoan \& Nordlund(2002)]{padoan02} Padoan, P., \&
Nordlund, {\AA}.\ 2002, \apj, 576, 870
\bibitem[Phan-Bao et 
al.(2003)]{phanbao03} Phan-Bao, N., et al.\ 2003, \aap, 401, 959
\bibitem[Phan-Bao et al.(2005)]{phanbao05} Phan-Bao, N., 
Mart{\'{\i}}n, E.~L., Reyl{\'e}, C., Forveille, T., 
\& Lim, J.\ 2005, Astronomische Nachrichten, 326, 1031 
\bibitem[Phan-Bao et al.(2006)]{phanbao06} Phan-Bao, N., 
Forveille, T., Mart{\'{\i}}n, E.~L., 
\& Delfosse, X.\ 2006, \apjl, 645, L153
\bibitem[Rayner et al.(2003)]{spex} Rayner, J.~T., Toomey, D.~W., Onaka, P.~M., Denault, A.~J., Stahlberger, W.~E., Vacca, W.~D.,
Cushing, M.~C., \& Wang, S.\ 2003, \pasp, 115, 362 %spex paper
\bibitem[Reipurth \& Clarke(2001)]{reipurth01} Reipurth, B., \&
Clarke, C.\ 2001, \aj, 122, 432
\bibitem[Scholz et al.(2006)]{scholz06} Scholz, A., 
Jayawardhana, R., \& Wood, K.\ 2006, \apj, 645, 1498
\bibitem[Seifahrt et 
al.(2005)]{seifahrt05} Seifahrt, A., Guenther, E., \& Neuh{\"a}user,
R.\ 2005, \aap, 440, 967 
\bibitem[Skrutskie et al.(2006)]{skrutskie06} Skrutskie, M.~F., et 
al.\ 2006, \aj, 131, 1163
\bibitem[Stamatellos et al.(2007)]{stamatellos07} Stamatellos, D., 
Hubber, D.~A., \& Whitworth, A.~P.\ 2007, \mnras, 382, L30
\bibitem[Tokovinin(1997)]{tokovinin97} Tokovinin, A.~A.\ 1997, \aaps,
  124, 75 
\bibitem[Vacca et al.(2003)]{spextool2} Vacca, W.~D., Cushing, M.~C.,
  \& Rayner, J.~T.\ 2003, \pasp, 115, 389
\bibitem[Vicente \& Alves(2005)]{vicente05} Vicente, S.~M., \& Alves,
  J.\ 2005, \aap, 441, 195
\bibitem[Reid \& Gizis(1997)]{reid97} Reid, I.~N., \& Gizis, J.~E.\ 1997, \aj, 114, 1992 
\bibitem[Reid et al.(2004)]{reid04} Reid, I.~N., et al.\ 2004, 
\aj, 128, 463
\bibitem[Weinberg et al.(1987)]{weinberg87} Weinberg, M.~D., 
Shapiro, S.~L., \& Wasserman, I.\ 1987, \apj, 312, 367 
\bibitem[Whitworth et al.(2007)]{whitworth07} Whitworth, A., Bate, 
M.~R., Nordlund, {\AA}., Reipurth, B., 
\& Zinnecker, H.\ 2007, Protostars and Planets V, 459

\end{thebibliography}
\end{document}